\documentstyle[aps,preprint]{revtex}

\newcommand{\bra}[1]{\mbox{$\langle #1 |$}}
\newcommand{\ket}[1]{\mbox{$| #1 \rangle$}}
\newcommand{\bracket}[2]{\mbox{$\langle {{#1}} \mathrel{ | {\vphantom
        {{#1} {#2}}} \kern-\nulldelimiterspace} {{#2}} \rangle$}}

\newcommand{\rem}[1]{}

\newcommand{\ros}{\rho_s}

\newcommand{\Hil}{\mbox{$\cal H$}}

\def\un{\leavevmode\hbox{\normalsize1\kern-4.6pt\large1}}

\begin{document}
\tightenlines 
\draft

\title {Quantum separability, time reversal and canonical decompositions}
\author{Anna Sanpera$^{\dag}$, Rolf Tarrach$^{\ddag}$ and Guifr\'e Vidal$^{\ddag}$}

\address{$^{\dag}$Centre d'Etudes de Saclay,  
Service des Photons, les Atomes et les Molecules/SPAM/DRCAM, 
\mbox{91191 Gif-Sur-Yvette,France.}}
\address{$^{\ddag}$ Departament d'Estructura i Constituents de la Materia,
Universitat de Barcelona, 
\mbox{08028 Barcelona, Espanya.}}

\date{\today}

\maketitle

\begin{abstract}
We propose an interpretation
of quantum separability based on a physical principle: local time
reversal. It immediately leads to a simple characterization
of separable quantum states that reproduces results known to hold for
binary composite systems and which thereby is complete for low dimensions. 
We then describe a constructive
algorithm for finding the canonical decomposition of separable and non
separable mixed states of dimensions 2x2 and 2x3. 

\end{abstract}

\narrowtext
\vspace{0mm}

Entanglement, inseparability and nonlocality are some of the most genuine
quantum concepts. As it has been pointed out, we still lack a complete
classification of quantum states in local and non-local ones\cite{pop1}, or more 
precisely we still lack the complete understanding of non-locality. Thus, while 
for pure states it is well established since long ago that the non-local character of the composite system is revealed in different but equivalent ways, the situation is drastically
different for mixed states.
For example, for pure states the violation of some kind of Bell inequalities\cite{bell}, or the demonstration that no local hidden variable models can account for the correlations between the observables in each subsystem, are equivalent definitions of non-locality. But for mixed states, described by density matrices, such equivalences fade away. Consider a composite quantum system described by a density
matrix $\rho$ in the Hilbert space ${\cal H}_a\otimes {\cal H}_b$. 
In the frame set by the concepts of our
starting sentence, product or factorizable states are the simplest possible.
They are of the form $\rho_p=\rho_a\otimes\rho_b$, i.e. for them, and only for them, the description of the two isolated subsystems is equivalent to the description of the 
composite system. Recalling that subsystems are described by the reduced
density matrices obtained via partial tracing: $\rho_a=Tr_b\rho$ (
$\rho_b=Tr_a\rho$), a density matrix corresponds to a product or factorizable state 
if and only if
\begin{equation}
\rho=Tr_b\rho\otimes Tr_a\rho  \iff \rho=\rho_p
\end{equation}
Also their index of correlation
defined in terms of von Neumann entropies of the system and subsystems,
\begin{equation}
I_c=Tr\rho \ln\rho-Tr \rho_a \ln\rho_a -Tr \rho_b\ln\rho_b
\end{equation}
vanishes, and this happens only for them\cite{bar}. Their subsystems are uncorrelated. Any state which is not a product state presents some kind
of correlation. They are called correlated states. 
Quantum mechanics has taught us that there is a hierarchy of 
correlations, and the physics in the different ranks is different.
The simplest correlated systems are the classically correlated or separable
systems. Their density matrices can always be written in the form:
\begin{equation}
\rho_s=\sum_{i} p_i \rho_{ai}\otimes\rho_{bi};\;\; 1\ge p_i\ge 0; \,\, \sum_ip_i=1
\label{separable}
\end{equation}
i.e. as a mixture of product states. 
Their characterization is notoriously
difficult. Thus, given a density matrix which is known to describe
a separable system no algorithm for decomposing it according to eq. (\ref{separable})
is known; besides, the decomposition is not unique.
In fact, only very recently Peres and the Horodecki family\cite
{per,ho3} have obtained a mathematical characterization of these states, at least when the dimension of the composite Hilbert space is $2\times 2$ or
$2\times 3$. For these cases the necessary and sufficient condition for separability is that the matrix obtained by partially transposing 
the density matrix $\rho$ is still a density matrix, i.e. hermitian, with unity  trace and non-negative eigenvalues
\begin{equation}
\rho^{T_b}=(\rho^{T_a})^*\ge 0 \iff \rho=\rho_s
\label{per}
\end{equation}
For composite systems described by Hilbert spaces of higher dimensions, the positivity condition of $\rho^{T_b}$ is only a necessary one for separability\cite{ho3}. 
Following the hierarchy of correlations, we find the states that are no longer separable (or classically correlated ), i.e.  $\rho\not=\rho_s$. These states are called  ``EPR-states''\cite{epr}, ``inseparable'', ``non-local'', and sometimes ``entangled'' or simply ``quantum-correlated'' to emphasize that  their correlations are not
strictly classical anymore, though often these labels do not refer to exactly the same states. This confusion reflects the need of a further subclassification of the inseparable states according to whether they admit local hidden variables, whether they violate some kind of Bell inequality\cite{wer,ho21}, etc..

The aim of this Letter is threefold. Firstly, we give a physical interpretation of the mathematical characterization of separability (Eq. (\ref{per})). Secondly, we provide a constructive ``canonical'' algorithm for decomposing any separable matrix (of dimension $\leq 6$ ) into a very small finite incoherent sum of product vectors. Finally, we show that a similar decomposition holds for inseparable states, with the signature of non-separability being expressed by some non-positive weights in the canonical decomposition. 

Let us first analyze the problem of separability from a physical point of view. We start by considering symmetry transformations in the Hilbert space of each subsystem. We limit ourselves in this paper to just binary composite
systems , i.e. ${\cal H}={\cal H}_a\otimes {\cal H}_b$. 
Wigner's theorem tells us that every symmetry transformation
should always be implemented by a unitary ($U$) or antiunitary ($A$) matrix. 
The direct product of unitary matrices $U_a\otimes U_b$ (or antiunitary matrices
$A_a\otimes A_b$) is a unitary (or antiunitary) matrix in the
Hilbert space of the composite system ${\cal H}$, and one can give an unambiguous definition of how such a transformation acts on any ket $\ket{\Psi}\in{\cal H}$. However, the combination of  a unitary and an antiunitary transformation $U_a\otimes A_b$ (or $A_a\otimes U_b$) 
results in  a transformation which is neither unitary nor antiunitary
in ${\cal H}$, whose action on a general ket of the composite
system  $\ket{\Psi}\in{\cal H}$, furthermore, cannot be properly defined. However, its action
on a product ket $\ket{e}\otimes\ket{f}\equiv\ket{e,f}$, (where $\ket{e}\in{\cal H}_a$ and
$\ket{f}\in{\cal H}_b$) is, but for a phase ambiguity, well defined. Thus, the
action of  combined transformation of the type $U_a\otimes A_b$ 
on projectors corresponding to pure product state is well defined without any ambiguity. 
As a separable state can always be
rewritten as a statistical mixture of product vectors:
\begin{equation}
\ros=\sum_i p_i (\ket{e_i}\bra{e_i}\otimes\ket{f_i}\bra{f_i});\,1\geq p_i\geq 0;\, \sum_ip_i=1
\label{decpro}
\end{equation}
it is clear that under the combined transformation $U_a\otimes A_b$ (or $A_a\otimes U_b$) $\rho_s$ transforms into:
\begin{equation}
\ros\rightarrow\rho^{'}_s=\sum_i p_i \left(\ket{e^{'}_i}\bra{e^{'}_i}\otimes\ket{f^{'}_i}\bra{f^{'}_i}\right) \,\,\,
\end{equation}
where $\ket{e^{'}_{i}}\equiv U_a\ket{e_i}\in{\cal H}_a$ ; $ \ket{f^{'}_i}\equiv A_b\ket{f_i}\in{\cal H}_b$. Therefore, $\ros^{'}$ describes also a physical state so that $\ros^{'}$ is a positive defined hermitian matrix (with normalized trace). 
This is what characterizes separable states: that any local symmetry transformation, which
obviously transforms local physical states into local physical states, also transforms the global physical state into another physical state. 
(Here and in what follows ``local'' means that it refers to the subsystems).

There exists only one independent antiunitary symmetry and its physical meaning is well known: time reversal. Any other antiunitary transformation can be
expressed in terms of time reversal (as the product of a unitary matrix times
time reversal). We are thus proposing that quantum separability of composite systems implies the lack of correlation between the time arrows of their subsystems. In other words: for separable states, the state one obtains by reversing time in one of its subsystems is also a physical state. Loosely speaking, systems which are classically correlated (separable) do not have memory of a unique time direction in the sense ``EPR'' correlated states have, and they are thus compatible with a time evolution which factorizes into the product of two opposed time evolutions.  Changing the time
direction in only one of the subsystems but not in both  
leads to a physical state since their time arrows are uncorrelated.

We can define now the ``separability'' operator as the simplest possible transformation of this type:

\begin{equation}
S\equiv I_a \otimes K_b
\label{timrev1}
\end{equation}
where $I_a$ stands for the identity acting in the first
subsystem  ${\cal H}_a$ and $K_b$ is the complex conjugation operator acting in ${\cal H}_b$\cite{gal}. It is straightforward to check that in the Hilbert-Schmidt basis\cite{ho22} of $2\times 2$ composite systems:
\begin{equation}
S\rho S = \rho ^{T_b}
\label{timrev2}
\end{equation}
for any $\rho$, whether it is separable or not, in spite of the fact that the action of $S$ on a general  $\ket{\Psi}\in{\cal H}$ cannot be properly defined. (This feature of being able to define a transformation on density matrices which one cannot define on kets is known for some nonunitary transformations, as e.g. a decohering time-evolution). Finally, as local time reversal is locally unitarily equivalent to expression (\ref{timrev1}), eqs.(\ref{per}) and (\ref{timrev2}) state that for 2x2 composite systems a state is separable if and only if it does not correlate local time flows. The same holds for a 2x3 composite system.
 
Let us go to the second point of our Letter, the ``canonical'' decomposition of a separable density matrix. Until very recently it was not known whether one could always find a finite incoherent sum of pure
product states for any separable $\rho_s$. P. Horodecki\cite{pawel} and Vedral et al.\cite{martin} 
have shown that any separable state $\rho_s$ can be 
decomposed into an incoherent sum of at most $N=(dim({\cal H}_a) \times dim({\cal H}_b))^2$ pure product states, although
no algorithm for obtaining this decomposition seems to be known. 
We will limit ourselves here, again, to the simplest possible case, i.e.
binary composite systems of dimensions $dim({\cal H}_a) = dim({\cal H}_b) = 2$ . In such cases, 
any separable density matrix can be written as 
a statistical mixture of at most $N$=16 pure product states.
Let us show here that one can do much better: indeed, any separable density matrix can be written as a convex combination
of at most $N=5$ pure product vectors. The whole proof of such a decomposition is based on the following theorems:

\noindent{\bf Theorem1} For any plane ${\cal P}_1$ in ${\cal C}^2\otimes {\cal C}^2$  defined by two product states $\ket{v_1}$ and $\ket{v_2}$
(where $\ket{v_i}=\ket{e_i}\otimes\ket{f_i}$; $\ket{e_i}\in{\cal H}_a$ and $\ket{f_i}\in{\cal H}_b$) either all the states in this plane are product states, or there is no other product states in it.\\
\noindent{\bf Theorem2} There exist planes ${\cal P}_2$ in ${\cal C}^2\otimes {\cal C}^2$ which contain only one product state.\\ 
\noindent{\bf Theorem 3} Any plane ${\cal P}_3$ in ${\cal C}^2\otimes {\cal C}^2$  contains at least one product state.

\noindent The proofs of the theorems are simple. It is convenient to express, with the help of the $SU(2)\otimes SU(2)$ transformations, the planes defined by the theorems (denoted as ${\cal P}_1$,${\cal P}_2$ and ${\cal P}_3$) as:
\begin{equation}
{\cal P}_1(\alpha_1,\beta_1)\equiv \alpha_1{1\choose 0}\otimes{1\choose 0} + \beta_1 {\cos A \choose e^{iB}\sin A}\otimes {\cos C \choose e^{iD}\sin C}
\label{teor1}
\end{equation} 
with $0\leq A,C\leq \pi/2$; $0\leq B,D<2\pi$, and $\alpha_1, \beta_1 \in {\cal{C}}$.
\begin{equation}
{\cal P}_2(\alpha_2,\beta_2)\equiv \alpha_2{1\choose 0}\otimes{1\choose 0} 
+ \beta_2 \pmatrix{ 0 \cr
                 cos A \cr
                 e^{iC}\sin A\cos B \cr
                 \sin A\sin B \cr}
\label{teor2}
\end{equation} 
with $0< A <\pi/2$; $0\leq B<\pi/2$; $0\leq C <2\pi$, and $\alpha_2, \beta_2 \in {\cal{C}}$. Finally:  
\begin{equation}
{\cal P}_3(\alpha_3,\beta_3)\equiv \alpha_3 
             \pmatrix{ \cos A \cr
                        0     \cr
                        0      \cr
                       \sin A   \cr} 
+ \beta_3 \pmatrix{ \sin A \cos B \cr
                 \sin B \cos C  \cr
                 e^{iD}\sin B\sin C \cr
                 -\cos A\cos B \cr}
\label{teor3}
\end{equation}
with $0< A <\pi/2$; $0\leq B,C\leq \pi/2$; $0\leq D <2\pi$, and $\alpha_3, \beta_3 \in {\cal{C}}$. There are further conditions that have to be imposed in  Eq.(\ref{teor3}) to ensure that the second state is not a product state. Solving for the values of $\alpha_i$ and $\beta_i$ allows to prove the theorems.
A consequence of the above theorems is the following corollary:\\

\noindent{\bf Corollary} If $\rho$ has rank 2 and is separable it can always
be written as a statistical mixture of two pure product states, and thus 
$\rho^{T_b}$ is also of rank 2.

Consider now a separable state $\ros$ such that both itself and its partially
transposed matrix are of rank 4,
\begin{equation}
r(\rho_s)=r(\rho_s^{T_b})=4
\label{ran4}
\end{equation}
(All other cases are subcases of this one, as we shall see immediately). Define now
\begin{equation}
\rho(p)\equiv \frac{1}{1-p}(\rho_s-p\ket{e_1,f_1}\bra{e_1,f_1}); \,\,\,
0 < p < 1
\label{p1}
\end{equation}
where $\ket{e_1}\in{\cal H}_a$ and $\ket{f_1}\in{\cal H}_b$  are completely arbitrary kets.
For $p$ small enough both $\rho$ and $\rho^{T_b}$  

\begin{equation}
\rho^{T_b}(p)\equiv \frac{1}{1-p}(\rho_s^{T_b}-p\ket{e_1,f_1^*}\bra{e_1,f_1^*});\,\,\,
0 < p < 1
\label{p1t}
\end{equation}
are positive, and therefore, due to eq.(\ref{per}), separable. Let us
denote by $p_1$ the smallest value for which a zero eigenvalue appears in $\rho(p)$
or $\rho^{T_b}(p)$. Let us assume that for $p_1$ one eigenvalue of $\rho(p)$ is equal to zero, i.e. $r(\rho(p_1))=3$ and 
$r(\rho^{T_b}(p_1))=4$ (the same argument holds for the opposite case).
Consider now a new pure product vector state belonging to the
range of $\rho(p_1)$; $\ket{e_2,f_2}\in{\cal R}(\rho(p_1))$
and define a new density matrix:
\begin{equation}
\bar\rho(p)\equiv \frac{1}{1-p}(\rho(p_1)-p\ket{e_2,f_2}\bra{e_2,f_2});\,\,\, 0 < p < 1.
\label{p2}
\end{equation}

\noindent As before, for $p$ small enough, both $\bar\rho(p)$ and $\bar\rho^{T_b}(p)$ are non-negative and
thus separable. Let us denote by $p_2$ the smallest value of $p$ for which either
$\bar\rho(p)$  or $\bar\rho^{T_b}(p)$ develop a new vanishing eigenvalue. It cannot be $ \bar\rho(p) $ unless, because of the corollary, $ \bar\rho^{T_b}(p)$  develops simultaneously two vanishing eigenvalues. Therefore, it is in general $\bar\rho^{T_b}(p)$ which will develop a new vanishing eigenvalue, so that 

\begin{equation}
r(\bar\rho(p_2))=r(\bar\rho^{T_b}(p_2))=3.
\label{ran3}
\end{equation}

As $\bar\rho(p_2)$ has a decomposition of the type of eq.(\ref{decpro}) 
with at least three terms, and $\bar\rho^{T_b}(p_2)$ has the corresponding partially transposed one,
there exist always a product state satisfying\cite{pawel} :
$\ket{e_3,f_3}\in{\cal R}(\bar\rho(p_2))$  and  $\ket{e_3,f_3^{\star}}\in{\cal R}(\bar\rho^{T_b}(p_2))$
where the following identity:
\begin{equation}
(\ket{\phi}\bra{\phi})^{T}=(\ket{\phi}\bra{\phi})^{\star}
\end{equation}
and some results by Hughston et al. \cite{hug} have been used. Define now:

\begin{equation}
\tilde\rho(p)\equiv \frac{1}{1-p}(\bar\rho(p_2)-p\ket{e_3,f_3}\bra{e_3,f_3});\,\,\, 
0 < p < 1.
\label{p3}
\end{equation}
\noindent It is clear from the above results that a $p_3$ exists such that:
\begin{equation}
r(\tilde\rho(p_3)\geq 0)=r(\tilde\rho^{T_b}(p_3)\geq 0)=2
\label{ran2}
\end{equation}
Finally, from the corollary, we know that there exists always the following
decomposition:

\begin{eqnarray}
\tilde\rho(p_3)&\equiv& p_4\ket{e_4,f_4}\bra{e_4,f_4}\nonumber\\
           &+& (1-p_4)\ket{e_5,f_5}\bra{e_5,f_5}; \,\,\, 0 < p_4 < 1.
\label{p45}
\end{eqnarray}
And from there our final results follows immediately:
\begin{eqnarray}
\ros&=&p_1P_1+p_2(1-p_1)P_2\nonumber \\
&+&p_3(1-p_2)(1-p_1)P_3\nonumber \\
&+&p_4(1-p_3)(1-p_2)(1-p_1)P_4\nonumber \\
&+&(1-p_4)(1-p_3)(1-p_2)(1-p_1)P_5
\label{p12345}
\end{eqnarray}
where $P_i\equiv\ket{e_i,f_i}\bra{e_i,f_i}$ are projectors onto pure product states.

It is obvious that this decomposition is far from unique. It is also clear
that often one can simplify the first two steps to one, i.e. go from 
expression (\ref{ran4}) to expression (\ref{ran3}) by subtracting only one single pure product state. This can always be done, by continuity, when for some $\ket{e,f}$ it happens that $r(\rho(p_1))=3$ and $r(\rho^{T_b}(p_1))=4$ and for others $r(\rho(p_1))=4$ and $r(\rho^{T_b}(p_1))=3$. We do not know, right now, whether it can always be done, i.e. if any separable density matrix can always be written
as a statistical mixture of just $N$=4 pure product states.

Let us now discuss our third result. Recall that for $dim({\cal H}) = 4$ a state 
has quantum correlations iff $\rho^{T_b}$ has at least one negative eigenvalue:

\begin{equation}
{\it inf}\, \sigma(\rho^{T_b})<0 \Longleftrightarrow \rho=\rho_q
\end{equation}
where $\sigma(\rho)$ means the spectrum of $\rho$ and the subscript $q$ means
quantum correlated. Let us prove now that there is only one negative
eigenvalue. If there were two, one could always find, according to our theorem,
a product state $\ket{e,f}$ in the plane defined by the
corresponding two eigenvectors, and for which obviously
\begin{equation}
\bra{e,f}\rho_q^{T_b}\ket{e,f}<0.
\end{equation}
But the above expression is equivalent, recalling the Hilbert-Schmidt decomposition, to
\begin{equation}
\bra{e,f^{\star}}\rho_q\ket{e,f^{\star}}<0,
\end{equation}
which is impossible. Let us then define

\begin{eqnarray}
\rho^{T_b}(p_1,p_2) &\equiv& \frac{1}{1+p_1+p_2} (\rho_q^{T_b} + p_1\ket{e_1,f_1}\bra{e_1,f_1}\nonumber \\
&+& p_2\ket{e_2,f_2}\bra{e_2,f_2})\;, p_i \geq 0,
\end{eqnarray}
where $\ket{e_i,f_i}$ are the two product states of the Schmidt decomposition of the negative eigenvalue eigenvector of $\rho_q^{T_b}$;
$\ket{n} = c_1\ket{e_1,f_1} + c_2\ket{e_2,f_2}$. For some finite values of $p_1$ and $p_2$, $\bar p_1$ and $\bar p_2$, $r(\rho_q^{T_b}(\bar p_1,\bar p_2) \geq 0) = 3$. As  $\rho_q(\bar p_1,\bar p_2) \geq 0$ the algorithm proceeds as before for the separable states. Thus
\begin{eqnarray}
\rho_q&=&(1+\bar p_1+\bar p_2) \rho_s(4)-\bar p_1 \ket{e_1,f_1^{\star}}\bra{e_1,f_1^{\star}}\nonumber \\
&-&\bar p_2 \ket{e_2,f_2^{\star}}\bra{e_2,f_2^{\star}},
\label{inse}
\end{eqnarray}
where $\rho_s(4)$ is a statistical mixture of $N=4$ pure product states. Certainly,
expression (\ref{inse}) is not a statistical mixture in the sense that two weights in 
the decomposition are negative, but that only means that $\rho_q$, which is a statistical mixture of pure states, is inseparable. 
Often one can find a decomposition of the type (\ref{inse}) but
with only five terms, either having $\rho_s(3)$ or only one $\bar p$, but we do not know
yet if this is always possible. Also, $\bar p_i$ are measures of the inseparability
of the state, which, supplemented with an adequate minimization procedure, might lead 
to a faithful quantification of entanglement \cite{bennett,lewen}.

Finally, let us illustrate our procedure with a simple example. Consider 
a pair of spin-$\frac{1}{2}$ particles in an impure state consisting of a singlet fraction $x$ and an isotropical mixture of the singlet and the triplet mixed in equal proportions\cite{wer}:

\begin{eqnarray}
\rho_w&=& {\it x}\ket{\Psi^{-}}\bra{\Psi^{-}}+\frac{(1-{\it x})}{4}
(\ket{\Psi^{-}}\bra{\Psi^{-}}\nonumber \\
&+& \ket{\Psi^{+}}\bra{\Psi^{+}}
+\ket{\Phi^{+}}\bra{\Phi^{+}}+\ket{\Phi^{-}}\bra{\Phi^{-}})
\label{wer}
\end{eqnarray}
where $0<x<1$,  $\ket{\Psi^{\pm}}\equiv 1/{\sqrt 2} (\ket{\uparrow\downarrow}\pm\ket{\downarrow\uparrow})$, and $\ket{\Phi^{\pm}}\equiv 1/{\sqrt 2}(\ket{\uparrow\uparrow}\pm\ket{\downarrow\downarrow})$.
The condition of separability (eq. \ref{per}) shows that $\rho_w$ is separable for $x\leq 1/3$ and inseparable otherwise.
One simple decomposition by following the procedure we have indicated is
given by:

\begin{equation}
\begin{array}{lll}
p_1=\frac{(1+3x)(1-x)}{4(1+x)}&;&\ket{e_1,f_1}={1\choose 0}\otimes {0\choose 1} \\
p_2=\frac{(1-3x)(1+x)^2}{(3+2x+3x^2)(1-x)}&;&\ket{e_2,f_2}={1\choose 0}\otimes {1\choose 0}\\
p_3=\frac{1}{3} &;&\ket{e_3,f_3}=\frac{1}{\sqrt{3x^2+1}}{2x\choose -\sqrt{1-x^2}}\otimes \frac{1}{\sqrt{2}}{\sqrt{1+x}\choose \sqrt{1-x}} \\
p_4=\frac{1}{2}&;&\ket{e_4,f_4}=\frac{1}{\sqrt{3x^2+1}}{2x\choose e^{i\pi/3}\sqrt{1-x^2}}\otimes \frac{1}{\sqrt{2}}{\sqrt{1+x}\choose e^{-i2\pi/3}\sqrt{1-x}}\\
 & &\ket{e_5,f_5}=\ket{e_4}^{\star}\otimes\ket{f_4}^{\star}\end{array}
\end{equation}
which also holds when $\rho_w$ is inseparable, in which case $p_2$ becomes negative. 
One can extend most of our results straightforwardly to $dim[\Hil]$=6, but not beyond, as Eq.(\ref{per}) does not hold anymore.
	
To summarize, we propose to characterize separable states by their inability of correlating local time arrows. 
For low enough dimensions this characterization is complete. We have also shown that it is always possible to find a decomposition
with at most five pure product states to express any separable density matrix (for $dim [{\cal H}] = 4$). Moreover, when the state is inseparable a similar decomposition with at most six pure product states holds, where
one or two of them have now a negative weight. We believe our results are
a step forward in the understanding of quantum (non)separability.

A.S. thanks  M. Lewenstein, P.  Horodecki and A. Peres for useful
discussions and acknowledges financial support of the European Community. R.T. enjoys financial support by CICYT (Spain), grant AEN95-0590 and by CIRIT (Catalonia),
grant GRQ93-1047. G.V. acknowledges a CIRIT grant.

\end{document}